\documentclass[letterpaper]{article}
\PassOptionsToPackage{hyphens}{url}
\usepackage{booktabs,diagbox,makecell,tabularx,multirow,enumerate,graphicx,url,tabularx}
\usepackage{dblfloatfix}

\begin{document}
\title{Whose Tweets are Surveilled for the Police:\\An Audit of a Social-Media Monitoring Tool via Log Files}

 \author{Glencora Borradaile\\School of Electrical Engineering and Computer Science\\Oregon State University
         \and Brett Burkhardt\\ School of Public Policy\\Oregon State University
         \and Alexandria LeClerc \\ School of Electrical Engineering and Computer Science\\Oregon State University}

\begin{abstract}
  Social media monitoring by law enforcement is becoming commonplace, but little is known about what software packages for it do.  Through public records requests, we obtained log files from the Corvallis (Oregon) Police Department's use of social media monitoring software called DigitalStakeout.  These log files include the results of proprietary searches by DigitalStakeout that were running over a period of 13 months and include 7240 social media posts.  In this paper, we focus on the Tweets logged in this data and consider the racial and ethnic identity (through manual coding) of the users that are therein flagged by DigitalStakeout.  We observe differences in the demographics of the users whose Tweets are flagged by DigitalStakeout compared to the demographics of the Twitter users in the region, however, our sample size is too small to determine significance.  Further, the demographics of the Twitter users in the region do not seem to reflect that of the residents of the region, with an apparent higher representation of Black and Hispanic people. We also reconstruct the keywords related to a Narcotics report set up by DigitalStakeout for the Corvallis Police Department and find that these keywords flag Tweets unrelated to narcotics or flag Tweets related to marijuana, a drug that is legal for recreational use in Oregon.  Almost all of the keywords have a common meaning unrelated to narcotics (e.g.\ broken, snow, hop, high) that call into question the utility that such a keyword based search could have to law enforcement.

  As social media monitoring is increasingly used for law enforcement purposes, racial biases in surveillance may contribute to existing racial disparities in law enforcement practices.  We are hopeful that log files obtainable through public records request will shed light on the operation of these surveillance tools.  There are challenges in auditing these tools: public records requests may go unfulfilled even if the data is available, social media platforms may not provide comparable data for comparison with surveillance data, demographics can be difficult to ascertain from social media and Institutional Review Boards may not understand how to weigh the ethical considerations involved in this type of research.  We include in this paper a discussion of our experience in navigating these issues.
\end{abstract}

\maketitle

\section{Introduction}

Law enforcement use of social media monitoring software has been in the news for several years, and usually it is not good news.  The ACLU of Northern California reported that MediaSonar, used by the Fresno Police Department, encouraged police to track \#BlackLivesMatter and related hashtags to identify ``threats to public safety''\cite{matt_cagle_this_2015}.  After it was revealed that MediaSonar marketed itself as a way for police to ``avoid the warrant process,'' Twitter cut off the company's access to their enterprise API\cite{mcquigge_experts_2017}.  Twitter also cut SnapTrends' API access after the release of details of law enforcement use of their software;  SnapTrends closed shop shortly thereafter\cite{dell_cameron_twitter_2016}.
Geofeedia was notably used during the Freddie Gray uprisings to ``arrest [protesters] directly from the crowd'' aided by social media posts and face recognition technology\cite{geofeedia_baltimore_2016}; shortly after this revelation from the ACLU of Northern California, Facebook, Twitter and Instagram all revoked API access from Geofeedia\cite{matt_cagle_facebook_2016}.
Both SnapTrends and Geofeedia are known to have enabled ``undercover'' accounts that befriend Facebook super-users in order to bypass users' privacy settings\cite{dell_cameron_twitter_2016}.  During a trial period of DigitalStakeout, an agent of the Oregon Department of Justice searched for \#BlackLivesMatter, discovered that an Oregon DOJ attorney was tweeting support and wrote a memo describing the posts as ``possible threats towards law enforcement'' -- the agent who wrote the memo was later found to be in violation of state law\cite{david_rogers_internal_2016}.

The usefulness of social media monitoring has been called into question.  Conarck reports that social media monitoring in Jacksonville, FL by Geofeedia ``included largely protected free-speech activity and useless miscellanea''\cite{conarck_sheriffs_2017}.  Relevant to the monitoring of social media in Corvallis, OR, in February 2018, an individual was arrested for Tweets threatening a shooting on the Oregon State University's Corvallis campus.  However, the Tweets were not discovered through surveillance of social media but through an anonymous tip line\cite{gazette-times_police_nodate}.  Indeed, our work echoes that of Conarck, uncovering that DigitalStakeout uses simple keyword search, at least on the topic of Narcotics, and that almost all the keywords have benign drug meanings that uncover ``useless miscellanea.''

Police increasingly utilize social media. A 2015 survey of over 500 US police departments found that 94\% of agencies had used social media in some capacity---to notify the public, recruit employees, gather intelligence, manage reputations, or other. The survey found that 89\% of agencies had used social media tools to further criminal investigations\cite{internationalassociationofchiefsofpoliceIACPSocialMedia2015}. Further, a 2016 report by the Brennan Center for Justice identified 151 local and state law enforcement agencies in the United States that have subscribed to social media monitoring services. These jurisdictions partner with a variety of private firms that deliver the monitoring service, including Geofeedia, Media Sonar, Snaptrends, Dataminr, DigitalStakeout, and Babel Street\cite{rachel_cohn_mapping_2016}.  What is known about social media monitoring technology is mostly gleaned from documents obtained through public records requests but these documents are often limited to marketing and training materials.  Meanwhile, the technologies are proprietary, and details of the underlying algorithms are unknown.

In this paper, we seek to understand how social media surveillance software may place certain groups of users under undue scrutiny.  Pew Research reports on the racial and ethnic, gender and age biases across the many social media platforms\cite{aaron_smith_social_2018}.  Sloan and Morgan report further demographic differences (in terms of gender, age, class and language) that exist among Twitter users as to whether they opt to geotag their tweets\cite{sloan_who_2015}.  We ask: Do these biases combine to unduly focus attention on certain users?  Does the software introduce biases that cannot be explained by a disparity in how different groups use social media?  We find that the demographics of the users whose Tweets are flagged by DigitalStakeout are representative of the demographics of the Twitter users in the region, but may not reflect that of the residents of the region, with an apparent higher representation of Black and Hispanic people.

To understand law enforcement monitoring of social media, we made public records requests to agencies asking for logs from social media monitoring tools.  We show that it is possible to reverse engineer the operation of keyword-based social media monitoring using log files.  We also show that we can audit the software\cite{sandvig_auditing_2014}, using the limited log files, for potential demographic disparities from the use of social media monitoring.  Because the data size is small and comes from a single jurisdiction, we are limited in the scope of questions we can answer.  However, this study provides a proof of concept and highlights areas for future study.

\subsection{Overview: From data to defining the research questions}

In the summer of 2017, we sent public records requests to 10 agencies listed by the Brennan Center as having (had) access to DigitalStakeout: Allentown Police Department, Alpharetta City Police Department, Corvallis Police Department, Fort Worth Police Department, Georgia Bureau of Investigation, Hillsboro County Sheriff's Office, Indiana State Police, Oregon State Police, Scottsdale Police Department, and Yakima Police Department.  We chose DigitalStakeout as a case study because it is a social media monitoring software package that was not reported to be subject to API restrictions by social media platforms (as MediaSonar, SnapTrends and Geofeedia were), is still actively used and had the largest number of listed subscribing agencies in the Brennan Center report.  Initially these requests were not made with a specific research question in mind, but more generally seeking to understand the use of social media monitoring software.  As part of the public records request, we asked for ``logs of searches that have been
input into DigitalStakeout'' and ``debug logs produced by DigitalStakeout.''

Several departments have claimed criminal investigatory material exemptions to public records laws (for which we are still seeking research exemptions to that exemption) and at least two agencies did not have records to release: Oregon DOJ did not subscribe to DigitalStakeout after their trial run (and now reports a policy of not subscribing to social media monitoring software) and the Yakima Police Department reports that their officers did not use the software and no longer subscribe.  The Corvallis Police Department did furnish logs in the form of .csv files which consist of 7240 links to social media posts, with some additional meta-data.  We describe the data in more detail in Section\ref{sec:data}.

Upon initial examination of the data, we observed that:  more people of color seemed to be represented in the collected social media posts than in Corvallis, and
the collected social media posts largely did not seem to be relevant to law enforcement.
These observations lead to the research questions:
\begin{enumerate}
\item Are the demographics of the social media users identified by DigitalStakeout representative of social media users or of the target population? (Section\ref{sec:demog})
\item How are the social media posts being identified by DigitalStakeout? (Section\ref{sec:keywords})
\end{enumerate}
At this point, we sought guidance from our IRB on how to responsibly pursue these questions.  We describe our procedure for demographic coding in Section\ref{sec:demo-code}.  An analysis of the racial and ethnic demographics are given in Section\ref{sec:demog}.  A look at the keywords used to flag social media posts is given in Section\ref{sec:keywords}. We describe our navigation of the ethical issues of this work in Section\ref{sec:ethics}.

\subsection{Related work}

While the extent of social media monitoring has been reported in the news, there is little work in the academic literature on the impacts of this.  The University of Chicago Crime Lab's report on using information gathered via social media to identify high school students for social service intervention is an exception, but it is not clear how they are monitoring social media\cite{the_university_of_chicago_crime_lab_connect_2019}.  As far as we know, no work in analyzing the actual tools used for monitoring social media has appeared in the academic literature.  The closest related work to ours is that which seeks to understand the algorithms and tools used for predictive policing, recidivism prediction, and face recognition.  We discuss work related to the demographic coding of social media users (which we do in this study) in Section\ref{sec:demo-code}.

Platforms for predictive policing may incorporate social media (such as Palantir\cite{brayne_big_2017}), but the impact of social media in predictive policing decisions has not been studied.  A simpler system for predictive policing, PredPol (whose basic algorithm is known and takes as input arrest data and reported incidents) has been the object of academic study.  Lum and Isaac demonstrate the existence and describe the potential consequences of feedback loops in PredPol\cite{lum_predict_2016} and Ensign etal.\ prove why these loops occur and suggest ways to prevent them\cite{ensign_runaway_2017}.

More closely related to our work is that of understanding COMPAS, a tool used to predict recidivism and used in parole decisions.  COMPAS hit the media after a ProPublica expose argued bias against black defendants\cite{julia_angwin_machine_2016}. ProPublica published their full data set of defendants, their demographics and abbreviated arrest history, and COMPAS scores, which they obtained through public records requests.  This data set allowed several academic teams to follow up with more in-depth statistical analyses and explanations (both supporting and criticizing the original analysis)\cite{chouldechova_fair_2017,dressel_accuracy_2018} and the development of theoretically-grounded models to explain and further understand the data\cite{adler_auditing_2018,friedler_impossibility_2016}.

Others have studied face recognition algorithms (which are increasingly being used in policing\cite{clare_garvie_perpetual_2016}) and have shown lower accuracies for younger, dark-skinned, or feminine faces\cite{klare_face_2012,buolamwini_gender_2018}.

Similar to these studies, we examine how software may introduce bias into law enforcement. However, whereas the previous studies relied on data derived from administrative records and physical appearance, the present study considers how social media users' online actions may expose them to more or less law enforcement attention. In particular, this study relies on public Twitter data. Such data are necessarily consciously curated by individual users, and the resulting public presentations may convey varying degrees of information. Thus, while social media monitoring presents one new method of surveilling citizens (perhaps differentially), it poses new challenges for how to define individual group membership and  subsequently measure aggregate levels of surveillance among different sub-populations.

\section{Description of the Data} \label{sec:data}

\begin{figure}[t]
  \centering
  \includegraphics[width=\columnwidth,keepaspectratio]{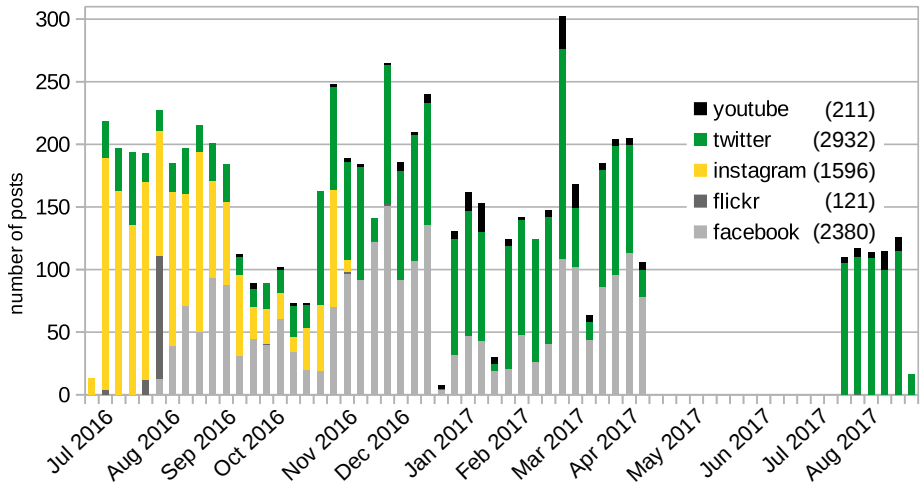}
  \vspace{-2em}
  \caption{Number of posts per week in the DigitalStakeout search logs from the Corvallis Police Department categorized by social media platform (total in brackets).}
  \label{fig:posts-by-platform}
\end{figure}

The {\em search logs} used in our audit contain the results of automated searches defined by DigitalStakeout (rather than the police department). They consist of 7240 links to social media posts, with some additional meta-data, over a period of 13 months (with a 3 month gap); see Figure\ref{fig:posts-by-platform}.  Also furnished by the Corvallis Police Department were additional {\em use logs} documenting officer-initiated inquiries in DigitalStakeout.  Our IRB denied our request to address research questions towards these additional use logs.  However, the use logs show that the Corvallis Police Department used DigitalStakeout infrequently and did not access the results of the automated searches defined by DigitalStakeout.

DigitalStakeout did not respond to our request for a demonstration, but the Corvallis Police Department did describe the system to us.  DigitalStakeout is provided as a subscription software that the Corvallis Police Department accesses through a web portal.  It provides three main ways to navigate social media, all within the predefined geographic region: (1) a map of the region with pins corresponding to recent posts of interest, (2) a search box for searching by name or screen name, and (3) an ``intelligence discovery'' tool that presents links referencing the geographic region of interest from the last hour.  A drop down menu gives access to posts captured by automated searches.  The use logs show that the Corvallis Police Department did not access the results of the automated searches.

As we see from Figure\ref{fig:posts-by-platform}, DigitalStakeout seems to inconsistently access all social media platforms except for Twitter.  From colleagues at the Brennan Center, we understand that the access to the Facebook API was pulled for all social-media monitoring platforms in Spring of 2017.

\subsection{Description of the search logs}
The data furnished by the Corvallis Police Department that we study here consists of 83 spreadsheets in comma-separated form, broken into 3 groups corresponding to different sets of search terms: LE (Law Enforcement) Terms, Terror Report, Narcotics.  The explanation accompanying the records request was that these are the results of search terms and (according to the Corvallis Police Department) ``all the search terms are preset proprietary lists of terms [DigitalStakeout] searches for.''  The columns of each spreadsheet include {\tt URL} and {\tt TIME}.  From July through early September, the Narcotics search logs include keywords for each social media post.  (We describe this in more detail in Section\ref{sec:keywords}.)

\begin{figure}[t]
  \centering
  \includegraphics[width=\columnwidth,keepaspectratio]{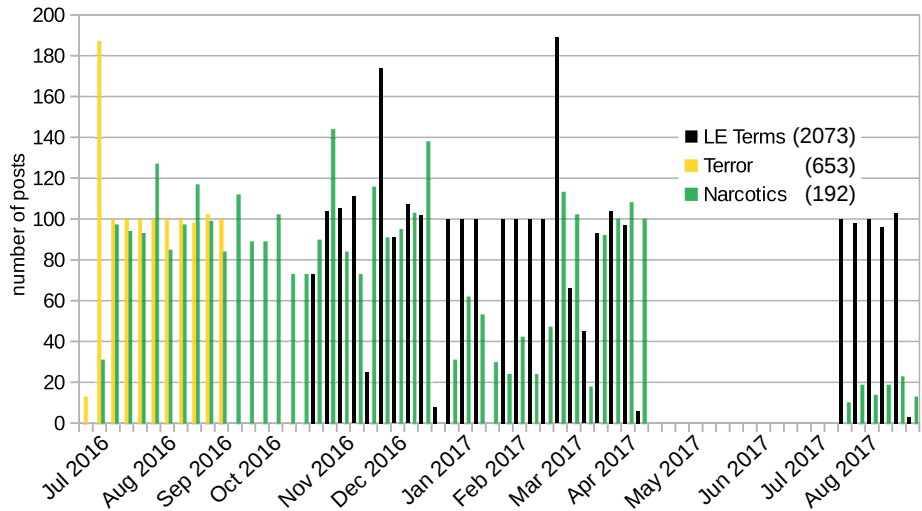}
  \vspace{-2em}
  \caption{Number of Tweets per week in the DigitalStakeout search logs from the Corvallis Police Department categorized by search term set (total in brackets).}
  \label{fig:posts-by-source}
\end{figure}

We note that for {\em LE Terms} and {\em Terror Report}, in many weeks exactly 100 search results are listed (Figure\ref{fig:posts-by-source}).  Indeed, in each of these weeks, the search logs seem to indicate that the search process collects social media posts until 100 results are returned then deactivates for the remainder of the week.

The vast majority of the social media posts are ``unprotected,'' arising from public accounts, so we believe DigitalStakeout to only be accessing publicly available data.  (We posit that those accounts that are currently not publicly available were made protected in the time since the posts were collected by DigitalStakeout.)  Herein, we focus on the subset of 2932 social media posts in the search logs that originate from Twitter, as Twitter's API allows us to sample comparative data sets (as we describe in Sections\ref{sec:CGT} and\ref{sec:hist}).  We only analyze data that is available on the date that we collect comparative samples (Table\ref{tab:data-size}).

\begin{table}[ht]
  \centering
  \begin{tabular}{lrrrr}\toprule
    && {\# tweets}  & & {\# users}  \\
    && {available} & & {available} \\
    &{\# tweets}& {07/12/19} &{\# users} & {05/31/18} \\ \midrule
    Terror & 192 & 101 & 86 & 74 \\
    Narcotics & 653 & 549 & 225 & 195 \\
    LE Terms & 2073 & 1442 & 774 & 595 \\ \bottomrule
  \end{tabular}
  \caption{Available data for analysis. We only use data that is publicly available on the date that we collect comparative samples.}
 \label{tab:data-size}
\end{table}

\subsection{Geography of the DigitalStakeout Tweets}

In conversation with the Corvallis Police Department, we learned that DigitalStakeout was calibrated to search social media posts originating from Benton County, Oregon (where Corvallis is located), but returned results outside of this county, notably from Benton County, Washington.

We examined the geotags of the Tweets and profile locations of the corresponding users.  All the Tweets in the Narcotics and Terror Report sets are geotagged with coordinates which lie within the 5 mile radius of Corvallis.  We presume that for these searches, DigitalStakeout is limiting the search of Twitter to Tweets with coordinate geotags in this region.  (Note that Tweets can also be tagged with a ``place'', which is a more general geographic region.)

On the other hand, the Tweets in the LE Terms data set seem to be collected with Twitter's ``place'' search.   The argument to a place search is either an ID or a place name.  A place ID is a precise identifier for a unique geographic region.  Given an ID, Twitter's place search will return Tweets: with geotags in that region; tagged with a place within that region; or, from a user with a profile location within that region if the Tweet has no place or coordinate geotag.  On the other hand, a place name is imprecise and Twitter will match that name to any geographic region that closely matches it.  It appears that the LE Terms search was configured with a descriptor relating to Benton County as the LE Terms data set contains Tweets with place and profile locations in Benton City, Washington and Bentonville, Arkansas.  Further, the LE Terms data set includes many ``retweets'', which in the past had geotags, but no longer do.  The current Twitter API will not pull retweets through a place search.  The poor configuration of the LE Terms search and change in the Twitter API has prevented us from being able to reproduce DigitalStakeout's geographical query for the LE Terms search.

\subsection{A comparative dataset of Corvallis Geotagging Tweeters} \label{sec:CGT}

To understand the demographics of the users whose Tweets appear in the DigitalStakeout data, we collect a sample of Tweets geotagged within the 5 mile radius of Corvallis, OR, matching the inferred geographical constraint for the Narcotics and Terror Report DigitalStakeout search results.  We were restricted to doing so with the public Twitter API, as our our interest in studying the demographics of the account users violates Twitter's Premium API agreements.\footnote{Twitter's Master License Agreement which one needs to sign to gain access to the Twitter Premium API ``may not be used [...] to target, segment, or profile individuals based on health (including pregnancy), negative financial status or condition, political affiliation or beliefs [...] racial or ethnic origin''.  Of course, Twitter's current Master License Agreement also seems to preclude social media monitoring itself: ``may not be used by [...] any public sector entity (or any entities providing services to such entities) for surveillance purposes, including but not limited to: (a) investigating or tracking Twitter’s users or their Content; and, (b) tracking, alerting, or other monitoring of sensitive events (including but not limited to protests, rallies, or community organizing meetings).''}

Twitter's public API geotag filter takes as input an input polygon and returns any Tweet whose geotag intersects that polygon.  The geotag of a Tweet is either a point (e.g. a GPS coordinate) or a polygon (rectangular, bounding a city, state or country, for example).  For our filter, we used the smallest rectangle encompassing Benton Co., OR.
We refiltered the collected tweets to those whose geotags were points within Corvallis's 5 mile radius, and matching the behavior of the Twitter Premium API geotag filter we infer was used to curate the DigitalStakeout data for the Narcotics and Terror Report searches.
We collected 1961 Tweets between March 6, 2018 and May 22, 2018.  From this, we sampled a set of 949 Twitter accounts to use in our comparative analysis\footnote{Initially we sampled 1000 accounts with Tweets within Benton Co.'s bounding box and readjusted the sample upon observing the different configuration for the LE Terms search.}; each account was selected for this final set with probability proportional to their frequency of Tweeting.   We call this set of accounts the Corvallis Geotagging Tweeters.  Of these 949 Twitter accounts, 102 accounts also appeared in the DigitalStakeout data set.

\section{Demographic coding} \label{sec:demo-code}

There is a body of literature on extracting demographic information from social media users in support of sociological and public health research that would be possible from the wealth of information available on social media platforms\cite{tyler_mccormick_using_2015}.  Cesare, Grant and Nsoesie discuss in detail many of the issues involved in inferring the demographic information from social media users, including the issue we dealt with most prominently: ``One challenge associated with the prediction of race and ethnicity is the need to create a clear, bounded definition. Racial and ethnic identity is complex and evaluations by others may not match an individuals’ self-identification.''\cite{cesare_detection_2017}

Cesare etal.\ also review 60 studies aimed at automatic detection of demographics, either using simple data detection (e.g.\ from profile descriptions) or matching (e.g.\ to user names) or machine learning techniques.  However, these techniques often limit the metadata they use to just profile photos or names in order predict demographics, and doing so limits the fraction of profiles for which demographics would be determinable.  For example, those methods that use profile photos to infer demographics only classify users with a profile photo containing a single face that they presume to be the profile owner.  In doing so, An and Weber discard 50\% of profiles\cite{jisun_an_greysanatomy_2016} and Messias, Vikatos, and Benevenuto discard 68\% of profiles\cite{messias_white_2017}.  In our case, since we are dealing with limited data and wish to avoid introducing any biases that may exist in users opting to use a profile photo that is of themself, these approaches are not appropriate.

The automatic detection methods that rely on machine learning techniques need a training set of data to seed the work.  In some cases this is generated from an external source (such MySpace's self-reported names and ethinicites\cite{jonathan_chang_epluribus:_2010} or mugshots.com arrest records\cite{shane_bergsma_broadly_2013}), or through using a secondary machine learning algorithm as a black box (such as Face++\cite{messias_white_2017}), or through manual coding much like we describe below.  While some groups perform in-house manual demographic coding similar to our own\cite{aron_culotta_predicting_2015,ehsan_mohammady_using_2014}, McCormick et al.recommend using Amazon Mechanical Turk for this task (and report on the reliability of doing this)\cite{tyler_mccormick_using_2015}.  However, since our IRB determined that our work was human subjects' research and requested a high level of data security, passing our data to MTurk workers would violate our approved protocols.

In the time since we completed our demographic coding, Preotiuc-Pietro and Unger presented a method which infers race and ethnicity from social media text and only require 100 posts from an account to predict demographics\cite{preotiuc-pietro_user-level_2018}.  Their model was robustly trained on a data set the authors built of users who self-report their race/ethnicity through a survey.  While the accuracy claims are quite strong, the authors have not responded to a request for access to their method.

\subsection{Protocol}

We coded all the DigitalStakeout accounts and Corvallis Geotagging Tweeters using the following protocol.  Before coding, we mixed the two data-sets, removed duplicates, and randomized the order of the accounts.  We did this for two reasons.  First, this would eliminate any bias that may be introduced from knowing that an account is or is not in the surveilled data.  Second, this provides some amount of privacy for the account holder from research scrutiny resulting from having been picked up by a surveillance tool.  We discuss this second point further in Section\ref{sec:ethics}.

We used the following publicly-available information to classify the gender and race of users with unprotected Twitter accounts:
\begin{itemize}
\item Name on the account (Twitter handle and profile name)
\item Profile and banner photo
\item Biography section (including links to external pages)
\item Recent tweets
\item Photos and videos available via the left sidebar
\end{itemize}
Using this information, coders first indicated whether the account belonged to an individual or an organization (e.g., company, band, school, group, etc.).

For individual accounts, coders classified the users' gender using the categories \{Female, Male, Other (for users who \textit{self-identify} as non-binary, gender fluid, transgender, genderqueer, or third gender), Don't know (if there is no image or text to indicate gender)\}.

Coders then classified the users' race using the categories: \{{\em White; Hispanic; Latino, or Spanish; Black or African American; Asian; American Indian or Alaska Native; Middle Eastern or North African; Native Hawaiian or Other Pacific Islander; Other (including users who self-identify as multiracial); Don't know (if there is no image or text to indicate race or ethnicity)}\}.

In our protocol, coder's looked first for positive evidence, such as self-identification, and then relied on photos or language in the absence of self-identification.

As shown below, several demographic categories appeared rarely, if at all, in the Twitter data. For the sake of more robust statistical comparisons, some analyses below collapse these race categories to, for example, \{{\em White; Black; Hispanic; Other; Don't Know}\}.

Gender and race are fluid and socially constructed categories, and there are other possible ways of categorizing the gender and race of users. However, we believe these categories provide a reasonable, though necessarily simplified, reflection of race and gender divisions in the US. Importantly, we determined that different coders following this protocol could reliably classify the race and gender of users.  Our protocol is similar to that used in other studies\cite{aron_culotta_predicting_2015,ehsan_mohammady_using_2014,tyler_mccormick_using_2015}.  We established the reliability of our coding protocol using multiple coders and measuring the inter-rater reliability, achieving  a substantial level of agreement using Krippendorff's alpha measure.  Details are in the appendix.

\section{Demographics: Race and Ethnicity}\label{sec:demog}

\begin{table}[ht]
  \begin{center}
    \begin{tabular}{lrrrr}\toprule
      & \makecell[c]{Corvallis} \\
      &\makecell[c]{Geotagging} & \multicolumn{3}{c}{DigitalStakeout} \\ \cmidrule(lr){3-5}
&	\makecell[c]{Tweeters}&	\makecell[c]{Narc.}&	\makecell[c]{Terr.}&	\makecell[c]{N+T}\\\midrule
$n$&	\makecell[c]{788}&	\makecell[c]{148}&	\makecell[c]{47}&	\makecell[c]{180}\\\midrule
White&	71.8\% &	78.4\% &	83.0\% &	78.9\%  \\
Black &6.5\% &	7.4\% &	4.3\% &	7.2\%  \\
    Hispanic&11.7\% &	7.4\% &	10.6\% &	7.8\% 	\\
    Other & 10.0\% & 6.8\%&	2.1\%&	6.1\% \\ \midrule
    $p$ & - & 0.23 & 0.25 & 0.13 \\\bottomrule
  \end{tabular}
\end{center}
Note: 15 users appear in both Narc and Terr.
  \caption{Coded Demographics---Narc. \& Terr.}
  \label{tab:cvi-twitter-dem}

\end{table}

\begin{table}\centering
  \begin{tabular}{lr}\toprule
    $n$ & 242\\ \midrule
White&	84.3\% \\
Black &4.1\% \\
    Hispanic&5.8\% \\
    Other & 6.0\% \\\bottomrule
  \end{tabular}
  \caption{Coded Demographics---LE Terms}
  \label{tab:le-terms-dem}
\end{table}

We report on our coded demographics for race and ethnicity for the Corvallis Geotagging Tweeters and those in the DigitalStakeout data who were coded as ``Individuals'' in our protocol.  We do not include users for whom there were neither images nor text to indicate race or ethnicity in these counts (which were coded ``Don't know'' according to our protocol).  In Table\ref{tab:cvi-twitter-dem}, we reduce the number of categories of race and ethnicity since the number of users coded in several categories were very small; in Table\ref{tab:cvi-twitter-dem}, ``Other'' encompasses several under-represented minorities: \{Asian, American Indian or Alaska Native, Middle Eastern or North African, Native Hawaiian or Other Pacific Islander, Other (including users who self-identify as multiracial)\}.  The full table of demographics is in Appendix\ref{sec:demo-race-all}.  We summarize gender demographics in Appendix\ref{sec:demo-gender}.

Users in the DigitalStakeout Narcotics and Terror Report data sets are drawn from Twitter in the same way as Corvallis Geotagging Tweeters.  We ask, are the users in the Narcotics and Terror Report data sets representative samples of Twitter users who geotag in Corvallis?  The p-values reported correspond to a Pearson Chi-squared test between CVI and Narc, Terr, N+T, respectively, for the race categories given in Table\ref{tab:cvi-twitter-dem}.  In each comparison, the race distributions differ, with white users appearing at higher rates in the DigitalStakeout sample than in the Corvallis Geotagging Tweeter sample. However, these differences are not statistically significant, a fact due in part to the small number of users in the DigitalStakeout samples.  We assume that the demographic distribution of geotagging Twitter users in Corvallis has not changed significantly from when the DigitalStakeout data was collected to when the Corvallis Geotagging Tweeters were sampled.

\begin{table}[b]\centering
  \begin{tabular}{lrrr}\toprule
    & only & \makecell{only or \\in combo.}& Hispanic \\\midrule
    White		&	83.8\% &87.5\%& 3.9\% \\
    Black &	1.1\%&1.8\%& 0.1\%\\
    Native American &	0.7\% &1.8\%& 0.2\%\\
    Asian		&	7.3\%&9.3\%& -\\
    Native HI/Pac.\ Isl'r	&		0.3\% &0.8\%& -\\
    some other race	&		2.8\% &3.2\%&2.6\%\\
    two or more races		&	4.0\% &N/A&0.6\%\\ \midrule
    Total &100.0\%&N/A&7.4\% \\
    \bottomrule
  \end{tabular}  \caption{Corvallis 2010 Census Demographics}
  \label{tab:census-dem}
\end{table}

\begin{table} [b]\centering
	\begin{tabular}{lr}
		\toprule
      $n$ (arrests) &  22,875 \\ \midrule
      White           &   89.4\% \\
		Black           &    3.5\% \\
		American Indian &    0.8\% \\
		Asian           &    1.7\% \\
		Unknown         &    4.6\% \\ \midrule
    Total &100.0 \\ \bottomrule
	\end{tabular}
	\caption{Benton County arrests, 2007-2012}
	\label{tab:bc-arrests}
\end{table}

As described above, the DigitalStakeout LE Terms data set seems to be drawn in a more general way that includes profile locations, and due to presumed poor configuration, includes users that seem to be from outside of Corvallis, OR.  Using the Twitter API, we examined the geotags of the Tweets in the LE Terms data set (if available) and user-described profile locations (as recorded on May 31, 2018 through the Twitter API).  We classified interpretable account profile locations according to whether they correspond to locations within Corvallis, OR or not.  (A profile location is non-interpretable if they did not correspond to mappable locations (such as {\em the moon} or {\em bliss}).)  The coded demographics of the accounts in the LE Terms data set that have Tweets geotagged or profile locations in Corvallis, OR are given in Table\ref{tab:le-terms-dem} (for the reduced set of categories -- full data given in Appendix\ref{sec:demo-race-all}).  These accounts would represent the same search, but configured for the Corvallis Police Department's region of interest.

\subsection{Demographics of the local population}

The demographics represented in Table\ref{tab:cvi-twitter-dem} are notably different from that of the Corvallis, OR (pop.\ 54,462) given in Table\ref{tab:census-dem}.  The census considers race orthogonal to ethnicity.  We give the fraction of the population that identifies as a {\em single} given race (column ``only''), the fraction of the population that identifies as a given race (alone or in combination with any other race), and the fraction of the population that identifies as Hispanic as well as any {\em single} given race.

Corvallis is also home to Oregon State University (OSU), with Spring 2018 enrollment of 28,568 students, 4,916 of which attended via e-campus alone.  OSU reports demographics of their domestic students, but not of the 11.72\% of the students who are international\cite{office_of_institutional_research_oregon_state_university_enrollment_2018}.  The demographics of the domestic students at OSU is similar to that of Corvallis demographics.

Arrests made by local police represent another relevant point of comparison, as they represent the population of local residents who are formally brought into the criminal justice system. According to data published by Lanfear\cite{lanfearExploringMentalHealth2013}, the demographics of arrests roughly mirror the demographics of the population. In particular, nearly 90\% of arrests made by Corvallis Police Department or Benton County Sheriff's Office from 2007-12 involved a White suspect (see Table\ref{tab:bc-arrests}).\footnote{The data omit Oregon State Police, which has jurisdiction over the Oregon State University campus.}

\subsection{Differences between Twitter users and the broader population} \label{ref:diff}

We wish to comment on the apparent difference in demographics between Twitter users and (Table\ref{tab:cvi-twitter-dem} and\ref{tab:le-terms-dem}) and Corvallis residents (Table\ref{tab:census-dem}).  It is impossible to determine the source of these differences, as there are many reasons to expect the differences we see in the demographics of these populations as (i) race is a significant factor for explaining difference in behavior, (ii) externally assigned (coded) demographics are a highly imperfect proxy for self-identified demographics, and (iii) the demographic categories we used for coding Twitter users are not perfectly comparable to Census categories.

To comment further on the first point, we refer to relevant literature and surveys which aim to quantify the demographic factors that play a role in social media use.

There are racial and ethnic differences in what social media platform people use.  For example: 28\% of Black U.S.\ adults use LinkedIn versus 13\% of Hispanic; 49\% of Hispanic U.S.\ adults use WhatsApp versus 14\% of White\cite{aaron_smith_social_2018}.  while Pew's report that 24\% of White people, 26\% of Black people, and 20\% of Hispanic people use Twitter does not explain the differences in the demographics of the populations we observe\cite{aaron_smith_social_2018}, there are two further considerations:  First, Pew's survey is nationwide, and there may be regional differences that compound racial and ethnic differences in social media use.  Second, Pew's survey does not drill down into how people interact with a given social media platform.  In particular, there could be racial and ethnic differences in whether people opt to geotag their Tweets.  Very few Tweets have geotags (measured at 0.85\% in 2013\cite{sloan_knowing_2013}), and Sloan and Morgan show that prevalence of geotagging varies among users depending on their gender, age, class and language\cite{sloan_who_2015}.

There are significant differences in Twitter use according to age: 45\% of 18-24 year-olds use Twitter compared to only 14\% of those over 50\cite{aaron_smith_social_2018}.  In a college town like Corvallis, this issue will be compounded.

Finally, we note that the Twitter users represented in Table\ref{tab:cvi-twitter-dem} are gathered purely based on geotags and that geotags will pick up Tweets from users who are simply visiting the area and not resident in Corvallis.

\section{Keywords} \label{sec:keywords}

In order to understand how the social media posts are being identified by DigitalStakeout, we attempt to reverse engineer the search.  We do so only for the ``Narcotics'' search.
Of the 101 Tweets that are still available (not deleted) in the ``Terror Report'' data set, 25 contain videos or images and 57 contain urls -- one Tweet contains only an image.  Given the limited and type of data and the likelihood that this search is not simply defined by a keyword search, we are not able to explain how the ``Terror Report'' is generated.  For ``LE Terms'', as previously noted, we are unable to reproduce the geographic filter.

The Narcotics dataset includes partial meta-data that suggests a simple keyword search is being employed: for the first 2 months of {\em Narcotics} search results, each social media post is accompanied by a set of keywords that match or closely match a word in the Tweet.  This seems to employ Twitter's keyword search which is more general than exact keyword matching: e.g., searching Twitter for ``hop'' will return Tweets containing the word ``hopped'' but not ``hope''.  We cluster keywords into keyword variant groups if they are variants of each other such as {\em rock, rocked, rocking, rocks}.  We use the simplest version in the group as a ``root'' representative (although it may not be a formal linguistic root).

There are 39 {\em known keywords} across 36 variant groups (listed in Table\ref{tab:nec}) that appear in the meta-data of the Narcotics dataset.  The known keywords explain 68\% of the available ``Narcotics'' Tweets.  We aim to uncover keywords that explain the remaining Tweets and develop a process that would reliably identify keywords should such meta-data not be available.

\subsection{A comparative dataset of historical Tweets} \label{sec:hist}

To understand how DigitalStakeout identifies Tweets in their Narcotics search, we collected the historical Tweets geotagged within the 5 mile radius of Corvallis over the same time period as the DigitalStakeout data using Twitter's Premium API.  The DigitalStakeout search logs suggest that there are time periods when the searches are not active in addition to the 3 month period for which we have no DigitalStakeout
data such a gaps in time between search log files.  To most conservatively represent the possible input accessed by DigitalStakeout, we down-sample the set of historical tweets to those with time-stamps between the first and last time stamps in a given search log for the Narcotics search.  This is imperfect, as there may be times during the creation of a search log in which the DigitalStakeout software is not active or in which the Twitter API is down.

\subsection{Reverse engineering keywords}

In order to reverse engineer the keywords used by DigitalStakeout, we compare the the (presumed) input to DigitalStakeout to the output from DigitalStakeout:
\begin{description}
\item[$T_{\mathrm{in}}$] The set of Tweets obtained with the same (presumed) geographical
  filter over the same period of time as the DigitalStakeout Narcotics search (as described in Section\ref{sec:hist}).
\item[$T_{\mathrm{out}}$] The set of Tweets in the DigitalStakeout Narcotics search log that are still publicly available.
\end{description}
Let $P$ be the set of words that appear only in DigitalStakeout Tweets; that is, words that appear in a Tweet of $T_{\mathrm{out}}$ but not in a Tweet of $T_{\mathrm{in}}\setminus T_{\mathrm{out}}$.  $P$ is the set of {\em possible} keywords.  If the search is active for all the periods of time that cover $T_{\mathrm{in}}$ and keywords are matched consistently, then a keyword must be in $P$.  Unfortunately, data is rarely {\em perfect}.  We find that 4 of the 39 known keywords (``yay'', ``broken'', ``trip'', ``tracks'') are not in $P$.  ``Broken'' is in 4 Tweets of $T_{\mathrm{in}} \setminus T_{\mathrm{out}}$; ``yay'', ``trip'', and ``tracks'' are each in 1 Tweet of $T_{\mathrm{in}} \setminus T_{\mathrm{out}}$.  This could be explained by the Twitter API or DigitalStakeout's services being down during this time period and not making data available for collection at the time of DigitalStakeout's collection.  That a Tweet with the word ``broken'' is missed 4 times is not surprising, as ``broken'' is overall a very high frequency word; indeed, ``broken'' is contained in 34\% of the available Narcotics Tweets.

Each Tweet in the Narcotics set contains at least one word of $P$.  For a Tweet in the Narcotics set that contains exactly one word $w$ from $P$, we presume that $w$ is the keyword that returned this Tweet.  We call the set of all such words in $P$ the set of {\em necessary} keywords and denote it $N$.  Given perfect data and exact keyword matching, all words in $N$ must be keywords.  For more general keyword matching but otherwise perfect data, all words in $N$ must be keywords (or variants of keywords).

We call the set of Tweets in the Narcotics set that do not contain a word or variant of a word in $N$ the set of {\em unexplained} Tweets.  Each Tweet in this set contains at least two words from $P$ and at least one word from $P \setminus N$.  Determining which words in $P\setminus N$ are keywords or derived from keywords is an impossible task.  The problem is a hitting set problem: find a subset $K$ of $P$ such that every Tweet contains a word of $K$.  There may be multiple feasible solutions, and no objective to decide between feasible solutions will necessarily correctly reverse engineer the set of keywords (or variants of keywords).  However, by examining the frequencies of words across the entire Narcotics dataset, we can determine a set of {\em likely} keywords: words that appear with higher frequency are more likely to have been keywords for the search.

We aim for an automated and reproducible method for identifying a set $L$ of likely keyword variant groups as follows.  For each unexplained tweet, let $W \in P\setminus N$ be a subset of words with a common root (a variant group) with highest frequency in the Narcotics dataset.  We let $L$ be the union of such variant groups that explain at least 2 DigitalStakeout Tweets ($T_{\mathrm{out}}$) that are not already explained by $N$.  We use a threshold of 2 to provide some confidence; a higher threshold could be used with a larger data set.

For the Narcotics data, $|P| = 607$, $|N| = 28$, and $|L| = 21$, where size measures the number of variant groups (e.g.\  {\em rock, rocked, rocking,} and {\em rocks} count as 1 variant group).  $N$ and $L$ explain 62\% of the Tweets in the Narcotics set.  We give the root form of the words in $N$ and $L$ in Table\ref{tab:nec} along with their frequency: the number of Tweets in $T_{\mathrm{out}}$ that these words (and their variants) appear in.  The full list of variants corresponding to these roots are given in the Appendix.

\begin{table*}[t]
  \centering
  \caption{Reverse-engineered and known keywords for the Narcotics search.}

  \label{tab:nec}
  \begin{tabular}[t]{lrlrlrlr}
    \multicolumn{8}{c}{Necessary}
  \\ \toprule
    Root & $f$ & Root & $f$  & Root & $f$ & Root & $f$\\ \midrule
{\bf snow}	&	54 & {\bf face}	&	11	&    {\bf trip}	&	4&{\bf waste}	&	2	\\
{\bf hop}	&	45	&{\bf cheese}	&	8&burger	&	4	&gang	&	2	\\
{\bf high}	&	40	&bag	&	8	&{\bf cook}	&	3	&hustle	&	2	\\
{\bf line}	&	22	&jack	&	7	&{\bf dope}	&	3	&rip	&	2	\\
{\bf party}	&	22	&treat	&	6	&{\bf blow}	&	3	\\
{\bf smoke}	&	14	&{\bf blast}	&	5&{\bf load}	&	3	\\
{\bf bowl}	&	13	&{\bf fried}	&	4&{\bf wreck}	&	3\\
    {\bf rock}	&	11	&{\bf crystal}	&	4&bake	&	3		\\	\bottomrule
  \end{tabular}\quad
    \begin{tabular}[t]{lrlrlr}
    \multicolumn{6}{c}{Likely}
  \\ \toprule
    Root & $f$ & Root & $f$  & Root & $f$ \\ \midrule
{\bf pie}	&	8	&      {\bf indica}	&	4&      growweed	&	2\\
{\bf pot}	&	6	&{\bf mash}	&	4	&{\bf burn}	&	2\\
{\bf zone}	&	6	&dank	&	4&keg	&	2\\
bud	&	6	&hip	&	4&malt	&	2\\
{\bf fade}	&	5&jam	&	4		&melt	&	2	\\
dabpro	&	5	&{\bf angel}	&	3\\
bang	&	5	&addict	&	3\\
deal	&	5	&roll	&	3\\ \bottomrule

\end{tabular}\quad  \begin{tabular}[t]{lr}
    \multicolumn{2}{c}{Missed Known}
  \\ \toprule
    Root & $f$ \\ \midrule
    {\bf broken} & 184\\
{\bf yay} & 3 \\
{\bf hookup} & 1 \\
{\bf stuck} & 1 \\
 {\bf munchies} & 1 \\
{\bf stash} & 0 \\
{\bf track} & 0 \\
                      {\bf tweed} & 0 \\ \bottomrule
                    \end{tabular}
\vspace{1em}\\
                    \begin{tabular}[t]{l}
Bold words are roots of known keywords. Frequency $f$ is the number of available Narcotics Tweets;
    $f= 0$ implies the \\corresponding Tweets are now deleted.  Missed Known words are those that were not discovered by reverse engineering.
                    \end{tabular}
\end{table*}

\subsection{Understanding the keywords}

Our method of reverse engineering is relatively robust for the following reasons:

\begin{itemize}
\item While $N$ and $L$ only explain 62\% of the Narcotics Tweets,
  $N \cup L \cup \{\mathrm{``broken''}\}$ explains 91\% of the
  Narcotics Tweets.  With a larger corpus of data, more
  noise-resistant methods would be able to capture missed words such
  as ``broken'' that were excluded as a possible keyword by relaxing
  the definition of {\em necessary}.

\item All but 3 of the keywords (keg, malt, melt) in $N$ and $L$ are drug terms, according to the DEA Drug Slang list\cite{noauthor_slang_2018} and the Urban Dictionary\footnote{\url{https://www.urbandictionary.com/}}.  Since our methods were oblivious to the meaning of the words, the words we uncover are quite likely to have been keywords.

\item $N$ and $L$ correctly identify 28 of 36 known keyword variant groups.  Of the remaining 8, 3 were not discoverable because they do not appear in any available Tweets, 3 have frequency 1 (so a lack of data make them difficult to discover) and the remaining 2 appeared in both $T_{\mathrm{in}}$ and $T_{\mathrm{out}}$ as discussed above.

\item Further of the words in $T_{\mathrm{out}}$ that do not appear in $N\cup L$, the only obvious drug term is ``marijuana'' which appeared in only one Tweet as part of a hashtag compounded with other words; this Tweet also contained a necessary keyword.
\end{itemize}

Note that all the keywords are English-language words or slang.  This may bias the search toward English-language users.  Among Twitter users geotagging in Corvallis, the effect is not large: 96.0\% of the Tweets in $T_{\mathrm{in}}$ and 98.9\% of the Tweets in $T_{\mathrm{out}}$ are labeled English-language by Twitter.\footnote{98.0\% of the Terror Report Tweets and 98.6\% of the LE Terms Tweets are labeled English-language by Twitter.}

Given these keywords, we feel that the Narcotics search results are unlikely to be useful for either risk assessment or sentiment analysis.  Of the 56 known, necessary and likely keyword variant groups, 15 are related to marijuana\cite{noauthor_slang_2018} and that recreational marijuana has been legal in Oregon for the entire period of DigitalStakeout data collection for the Corvallis Police Department.  Many of the Tweets containing marijuana-related keywords are from one of the many state-regulated marijuana dispenseries in Corvallis.  Many of the keywords, although drug-related, are sufficiently general that they pick a lot of Tweets that are unrelated to narcotics.  For example, ``broken'' picks up 178 Tweets from a weather bot that reports ``broken clouds'' as the forecast (and only 6 other Tweets).  Variants of the word ``hop'' (which pertain to drug use) exclusively pick up Tweets from local breweries (of which Corvallis is home to many).  Likewise ``bowl'' exclusively flags Tweets about the game of bowling or bowls of food.  Finally, ``party'' picks up the variants ``\#kidsparty'', ``\#pizzaparty'', and ``\#birthdayparty'' which are unlikely to be related to narcotics.

\section{Research ethics}\label{sec:ethics}

When we initially made our public records requests we didn't know if we would receive any data, never mind what form that data would take, or whether it would include personally identifiable information (PII).   Upon receiving the data (which includes PII in the form of links to social media posts that are created by individuals, many of whom associate their account with their real identity) and  formulating our research questions, we embarked on gaining IRB approval for our research.  We did so prior to pulling comparative data sets through the Twitter API.

Discussions with colleagues regarding this work received mixed opinions as to whether this research is Human Subjects Research at all, with opinions seeming to fall along disciplinary lines.  In fact, our IRB took one full-board meeting to decide that question alone.  In deciding whether the proposed work falls under Human Research Protection, consider the Office for Human Research Protections' guidelines for deciding ``Is an Activity Research Involving Human Subjects?''\footnote{\url{https://www.hhs.gov/ohrp/regulations-and-policy/decision-charts/index.html\#c1}}  Although our work does not involve intervention or interaction with individuals, the information does include PII.  Deciding whether this research is Human Subjects Research thus comes down to deciding if this information is {\em private}: ``About behavior that occurs in a context in which an individual can reasonably expect that no observation or recording is taking place, or provided for specific purposes by an individual and which the individual can reasonably expect will not be made public.''

Although by the letter of the law, the data we used (both collected from Twitter and obtained from the Corvallis Police Department) is public (or was at the time of collection) and so may be considered exempt from IRB oversight, one also needs to consider whether someone would expect their data (and the association of their data with a given commercial/state surveillance dataset) to be used for research.  While many people are aware of the extent of digital surveillance\cite{rainie_how_2015}, few would expect their public social media posts to be collected by private company, furnished to a police department and logged, made the subject of a public records request to finally end up in the hands of a researcher.  Indeed, Fiesler and Proferes report that ``few [Twitter] users were previously aware that their public tweets could be used by researchers, and the majority felt that researchers should not be able to use tweets without consent.''\cite{fiesler_participant_2018}

Along these lines, our IRB determined that our research is Human Subjects Research.  The permission to pursue our research is under expedited categories 5 \& 7 (minimal risk to adults and minimal risk to children -- \S 46.404). We obtained a waiver of informed consent, but only by agreeing to the highest level of data protections.

\subsection{Waiver of informed consent}

We argue that the importance of shedding light on proprietary software being used for policing outweighs the risk to an individual user that may (unknowingly) participate in this research.  We also argue that this research would not be possible if informed consent was required.  Not only would it likely reduce the available data significantly and introduce more biases, the very act of seeking out consent from people whose Tweets were collected by DigitalStakeout makes a data breach much more likely.
We do respect users explicit choices by ignoring accounts marked as protected although these accounts still display name, profile location, profile banner and profile photos.

One cannot forget that Twitter and other social media users are not informed that they are being monitored by DigitalStakeout -- it is far from common knowledge that the Corvallis Police Department subscribes to DigitalStakeout.  Publication of this and similar research serves as one means of communicating this fact.

\subsection{Data protections}

In order to obtain a waiver of consent, in balancing risk and benefit, our IRB required the highest level of data protections.  This involves storing information in a manner that provides access only to authorized individuals.  Our data is stored on encrypted drives and shared between the study staff via end-to-end encrypted channels.  Prior to demographic coding, DigitalStakeout data was randomly mixed with the Corvallis Geotagging Tweeters data.  The demographically coded PII (social media user names) was kept separate from indicators as to their source.  All this was to minimize the risk of leaking that a given user was included in a surveillance data set.  However, this does preclude sharing the data more broadly for further study: the demographic data with search type (e.g. Narcotics) but without PII (the social media user name) may be enough information to de-anonymize users in the small community of Corvallis, much like the famed ZIP, gender, DOB de-anonymizing observation of Sweeney\cite{sweeney_simple_2000}.  Of course, another researcher could request the same information from the Corvallis Police Department as we did.  Alternatively, we will work with other researchers in collaboration with our IRB to ensure that further research can be pursued.  As part of our currently approved IRB research protocols, we can only share aggregate data and keywords, but not precise Tweets.  Although the keywords could be used to reproduce DigitalStakeout ``Narcotics'' searches, since the DigitalStakeout searches were not running continuously and we are not publishing precise times during which the searches were active, it is impossible to perfectly reproduce the output of the DigitalStakeout search logs.

\section{Conclusion} \label{sec:fw}

This study has shown that we can indeed use log files to audit social media monitoring software and address our research questions.  First, we find small but non-significant differences in the race distribution of users flagged by DigitalStakeout and users geotagging their Tweets in Corvallis. Second, we were able to, for the Narcotics search, able to reverse engineer the keywords most likely used and show that this method is robust by comparing to a subset of keywords available through the metadata.

Racial disparities exist throughout the justice system, including in policing. Research has
found that non-whites are more likely than whites to be arrested or stopped, net of legally
relevant factors like crime type and presence of witnesses\cite{gelmanAnalysisNewYork2007,Beckett2006,eppPulledHowPolice2014,Kochel2011}. These disparities in police contact contribute to a severe
overrepresentation of people of color in US prisons \cite{Alexander2010,Western2006}.  Given this, we argue that it is important to be able to audit tools used in the justice system (such as has been done for recidivism prediction, face recognition, and predictive policing) for racial disparities.  Social media monitoring is simply another avenue for creating disparities, and there are many points at which an inequity could be introduced: including access to social media, adoption of a particular social media platform, interacting with the platform in a way that gives access to monitoring software, and using certain keywords. We have shown that geotagging or setting a profile location are choices that result in access by DigitalStakeout for monitoring.  Others have shown that such choices are likely to correlate with demographics\cite{sloan_knowing_2013}.  We have also shown that many keywords flag benign Tweets, at least from a risk-assessment perspective; this could draw undue attention from law enforcement.

Whether the purpose of social media monitoring by police is for sentiment analysis or risk assessment, unless the population that is affected mirrors that of the police jurisdiction, the bias will result in a skewed view of the population (if used for sentiment analysis) or undue attention on one subpopulation over another (in the case of risk assessment).

The use of log files is useful in gaining insight into the proprietary tools.  We would recommend that log files be required and available for research or other independent evaluation to ensure transparency of the algorithms that are reshaping law enforcement.  Policy could help overcome the difficulties of this audit, including lack of data, poorly or incompletely logged data and inaccessibility of data (from all entities involved, including law enforcement agencies, social media monitoring software houses and social media platforms).

\paragraph{Acknowledgements}  We would like to acknowledge the help of Alexander Guyer and Baigong Zheng with data collection and generation efforts and help from the Civil Liberties Defense Center with pursuing public records requests.  We would also like to acknowledge the funding support of the College of Engineering Dean's Professorship.
\newpage

\bibliographystyle{plain}
\bibliography{borradaile}

\begin{thebibliography}{10}

\bibitem{aaron_smith_social_2018}
{Aaron Smith} and {Monica Anderson}.
\newblock Social {Media} {Use} in 2018.
\newblock Technical report, Pew Research Center, March 2018.

\bibitem{adler_auditing_2018}
Philip Adler, Casey Falk, Sorelle Friedler, Tionney Nix, Gabriel Rybeck, Carlos
  Scheidegger, Brandon Smith, and Suresh Venkatasubramanian.
\newblock Auditing black-box models for indirect influence.
\newblock {\em Knowledge and Information Systems}, 54(1):95--122, January 2018.

\bibitem{noauthor_slang_2018}
Drug~Enforcement Administration.
\newblock Slang {Terms} and {Code} {Words}: {A} {Reference} for {Law}
  {Enforcement} {Personnel}.
\newblock {DEA} {Intelligence} {Report} DEA PRB 06-13-18-25, Drug Enforcement
  Administration, July 2018.

\bibitem{Alexander2010}
Michelle Alexander.
\newblock {\em The New {{Jim Crow}}: Mass Incarceration in the Age of
  Colorblindness}.
\newblock {New Press}, {New York}, 2010.

\bibitem{aron_culotta_predicting_2015}
{Aron Culotta}, {Nirmal Kumar Ravi}, and {Jennifer Cutler}.
\newblock Predicting the {Demographics} of {Twitter} {Users} from {Website}
  {Traffic} {Data}.
\newblock In {\em Proceedings of the {Twenty}-{Ninth} {AAAI} {Conference} on
  {Artificial} {Intelligence}}, pages 72--78, Austin, TX, 2015. Association for
  the Advancement of Artificial Intelligence.

\bibitem{Beckett2006}
Katherine Beckett, Kris Nyrop, and Lori Pfingst.
\newblock Race, {{Drugs}}, {{And Policing}}: {{Understanding Disparities In
  Drug Delivery Arrests}}.
\newblock {\em Criminology}, 44(1):105--137, 2006.

\bibitem{brayne_big_2017}
Sarah Brayne.
\newblock Big data surveillance: {The} case of policing.
\newblock {\em American Sociological Review}, 82(5):977--1008, 2017.

\bibitem{buolamwini_gender_2018}
Joy Buolamwini and Timnit Gebru.
\newblock Gender {Shades}: {Intersectional} {Accuracy} {Disparities} in
  {Commercial} {Gender} {Classification}.
\newblock In {\em Proceedings of {Machine} {Learning} {Research}}, volume~81 of
  {\em Conference on {Fairness}, {Accountability} and {Transparency}}, pages
  1--15, New York University, NYC, January 2018. ACM.

\bibitem{cesare_detection_2017}
Nina Cesare, Christan Grant, and Elaine Nsoesie.
\newblock Detection of {User} {Demographics} on {Social} {Media}: {A} {Review}
  of {Methods} and {Recommendations} for {Best} {Practices}.
\newblock Technical Report 1702.01807, arXiv, 2017.

\bibitem{chouldechova_fair_2017}
Alexandra Chouldechova.
\newblock Fair {Prediction} with {Disparate} {Impact}: {A} {Study} of {Bias} in
  {Recidivism} {Prediction} {Instruments}.
\newblock {\em Big Data}, 5(2):153--163, June 2017.

\bibitem{clare_garvie_perpetual_2016}
{Clare Garvie}, {Alvaro Bedoya}, and {Jonathan Frankle}.
\newblock The {Perpetual} {Line}-{Up}: {Unregulated} {Police} {Face}
  {Recognition} in {America}.
\newblock Technical report, Georgetown Law, Center on Privacy \& Technology,
  2016.

\bibitem{conarck_sheriffs_2017}
Ben Conarck.
\newblock Sheriff's {Office}'s social media tool regularly yielded false
  alarms.
\newblock {\em The Florida Times-Union}, May 2017.

\bibitem{daniel_klein_kappaetc:_2016}
{Daniel Klein}.
\newblock {KAPPAETC}: {Stata} module to evaluate interrater agreement.
\newblock Statistical {Software} {Components} {S}458283, Boston College
  Department of Economics, revised 01 Feb 2018, 2016.

\bibitem{david_rogers_internal_2016}
{David Rogers}.
\newblock An {Internal} {Report} on {Oregon}'s {Illegal} {Surveillance} of
  {Black} {Lives} {Matter} on {Twitter} {Leaves} {Us} {With} {More} {Questions}
  {Than} {Answers}.
\newblock Technical report, American Civil Liberties Union, April 2016.

\bibitem{dell_cameron_twitter_2016}
{Dell Cameron}.
\newblock Twitter {Cuts} {Ties} with {SnapTrends}, a {Social} {Media} {Spying}
  {Tool} for {Police}.
\newblock Blog, The Daily Dot, October 2016.

\bibitem{dressel_accuracy_2018}
Julia Dressel and Hany Farid.
\newblock The accuracy, fairness, and limits of predicting recidivism.
\newblock {\em Science Advances}, 4(1):1--5, January 2018.

\bibitem{ehsan_mohammady_using_2014}
{Ehsan Mohammady} and {Aron Culotta}.
\newblock Using {County} {Demographics} to {Infer} {Attributes} of {Twitter}
  {Users}.
\newblock In {\em Proc. of {Workshop} on {Social} {Dynamics} and {Personal}
  {Attributes} in {Social} {Media} {Proceedings} of the {Workshop}}, pages
  7--17, Baltimore, Maryland, 2014. Association for Computational Linguistics.

\bibitem{ensign_runaway_2017}
Danielle Ensign, Sorelle Friedler, Scott Neville, Carlos Scheidegger, and
  Suresh Venkatasubramanian.
\newblock Runaway {Feedback} {Loops} in {Predictive} {Policing}.
\newblock In {\em Proc. of {Fairness}, {Accountability}, and {Transparency} in
  {Machine} {Learning} {Workshop}}, volume~81, pages 1--12, New York
  University, NYC, 2017. ACM.

\bibitem{eppPulledHowPolice2014}
Charles~R. Epp, Steven {Maynard-Moody}, and Donald {Haider-Markel}.
\newblock {\em Pulled over: {{How}} Police Stops Define Race and Citizenship}.
\newblock {University of Chicago Press}, {Chicago}, 2014.

\bibitem{fiesler_participant_2018}
Casey Fiesler and Nicholas Proferes.
\newblock "{Participant}" {Perceptions} of {Twitter} {Research} {Ethics}.
\newblock {\em Social Media + Society}, 4(1):205630511876336, January 2018.

\bibitem{friedler_impossibility_2016}
Sorelle Friedler, Carlos Scheidegger, and Scheidegger~Suresh
  Venkatasubramanian.
\newblock On the (im)possibility of fairness.
\newblock September 2016.

\bibitem{gelmanAnalysisNewYork2007}
Andrew Gelman, Jeffrey Fagan, and Alex Kiss.
\newblock An {{Analysis}} of the {{New York City Police Department}}'s
  ``{{Stop}}-and-{{Frisk}}'' {{Policy}} in the {{Context}} of {{Claims}} of
  {{Racial Bias}}.
\newblock {\em Journal of the American Statistical Association},
  102(479):813--823, September 2007.

\bibitem{geofeedia_baltimore_2016}
{Geofeedia}.
\newblock Baltimore {County} {Police} {Department} and {Geofeedia} {Partner} to
  {Protect} the {Public} {During} {Freddie} {Gray} {Riots}, October 2016.

\bibitem{internationalassociationofchiefsofpoliceIACPSocialMedia2015}
{International Association of Chiefs of Police}.
\newblock {{IACP Social Media}}: {{Publications}}.
\newblock http://www.iacpsocialmedia.org/resources/publications/, 2015.

\bibitem{jisun_an_greysanatomy_2016}
{Jisun An} and {Ingmar Weber}.
\newblock \#greysanatomy vs. \#yankees: {Demographics} and {Hashtag} {Use} on
  {Twitter}.
\newblock In {\em Proc. of the 10th {International} {Conference} on {Weblogs}
  and {Social} {Media} ({IWCSM})}, pages 523--526, Cologne, Germany, 2016.
  AAAI.

\bibitem{jonathan_chang_epluribus:_2010}
{Jonathan Chang}, {Itamar Rosenn}, {Lars Backstrom}, and {Cameron Marlow}.
\newblock epluribus: {Ethnicity} on {Social} networks.
\newblock In {\em Proc. of the {Fourth} {International} {Conference} on
  {Weblogs} and {Social} {Media} ({ICWSM})}, pages 18--25, Washington, DC,
  2010. AAAI.

\bibitem{julia_angwin_machine_2016}
{Julia Angwin}, {Jeff Larson}, {Surya Mattu}, and {Lauren Kirchner}.
\newblock Machine {Bias}: {There}'s software used across the country to predict
  future criminals. {And} it’s biased against blacks.
\newblock Report, ProPublica, May 2016.

\bibitem{kilem_gwet_handbook_2014}
{Kilem Gwet}.
\newblock {\em Handbook of {Inter}-{Rater} {Reliability}}.
\newblock Advanced Analytics Press, Maryland, USA, 2014.

\bibitem{klare_face_2012}
Brendan Klare, Mark Burge, Joshua Klontz, Richard Bruegge, and Anil Jain.
\newblock Face {Recognition} {Performance}: {Role} of {Demographic}
  {Information}.
\newblock {\em IEEE Transactions on Information Forensics and Security},
  7(6):1789--1801, December 2012.

\bibitem{Kochel2011}
Tammy~Rinehart Kochel, David~B. Wilson, and Stephen~D. Mastrofski.
\newblock Effect {{Of Suspect Race On Officers}}' {{Arrest Decisions}}.
\newblock {\em Criminology}, 49(2):473--512, 2011.

\bibitem{landis_measurement_1977}
J.~Richard Landis and Gary Koch.
\newblock The measurement of observer agreement for categorical data.
\newblock {\em Biometrics}, 33(1):159--174, March 1977.

\bibitem{lanfearExploringMentalHealth2013}
Charles~C. Lanfear.
\newblock {\em Exploring a {{Mental Health Crisis}} : {{An Examination}} of
  {{Mental Health Arrests}} in {{Benton County}}, {{OR}}}.
\newblock M.{{S}}. {{Thesis}}, Oregon State University, {Corvallis, OR}, 2013.

\bibitem{lum_predict_2016}
Kristian Lum and William Isaac.
\newblock To predict and serve?
\newblock {\em Significance}, 13(5):14--19, October 2016.

\bibitem{matt_cagle_this_2015}
{Matt Cagle}.
\newblock This {Surveillance} {Software} is {Probably} {Spying} on
  \#{BlackLivesMatter}.
\newblock Report, ACLU of Northern CA, December 2015.

\bibitem{matt_cagle_facebook_2016}
{Matt Cagle}.
\newblock Facebook, {Instagram}, and {Twitter} {Provided} {Data} {Access} for a
  {Surveillance} {Product} {Marketed} to {Target} {Activists} of {Color}.
\newblock Report, ACLU of Northern CA, October 2016.

\bibitem{mcquigge_experts_2017}
Michelle McQuigge.
\newblock Experts divided on social media surveillance after {Twitter} pulls
  plug on {Media} {Sonar}.
\newblock {\em The Hamilton Spectator}, January 2017.

\bibitem{messias_white_2017}
Johnnatan Messias, Pantelis Vikatos, and Fabrício Benevenuto.
\newblock White, man, and highly followed: {Gender} and race inequalities in
  {Twitter}.
\newblock In {\em Proc. of the {International} {Conference} on {Web}
  {Intelligence} ({WI}'17)}, pages 1--9, Leipzig, Germany, August 2017.
  IEEE/ACM.

\bibitem{office_of_institutional_research_oregon_state_university_enrollment_2018}
{Office of Institutional Research, Oregon State University}.
\newblock Enrollment {Summary} - {Spring} {Term}, April 2018.

\bibitem{preotiuc-pietro_user-level_2018}
Daniel Preotiuc-Pietro and Lyle Ungar.
\newblock User-{Level} {Race} and {Ethnicity} {Predictors} from {Twitter}
  {Text}.
\newblock In {\em Proceedings of the International Conference on Computational
  Linguistics}, page~12, 2018.

\bibitem{rachel_cohn_mapping_2016}
{Rachel Cohn} and {Angie Liao}.
\newblock Mapping {Reveals} {Rising} {Use} of {Social} {Media} {Monitoring}
  {Tools} by {Cities} {Nationwide}.
\newblock Report, Brennan Center for Justice, November 2016.

\bibitem{rainie_how_2015}
Lee Rainie and Mary Madden.
\newblock How {People} are {Changing} {Their} {Own} {Behavior}.
\newblock Technical report, Pew Research Center, March 2015.

\bibitem{sandvig_auditing_2014}
Christian Sandvig, Kevin Hamilton, Karrie Karahalios, and Cedric Langbort.
\newblock Auditing {Algorithms}: {Research} {Methods} for {Detecting}
  {Discrimination} on {Internet} {Platforms}.
\newblock In {\em Proc. of 64th {Annual} {Meeting} of the {International}
  {Communication} {Association} - {Data} and {Discrimination}: {Converting}
  {Critical} {Concerns} into {Productive} {Inquiry}}, page~23, Seattle, WA, May
  2014. ICA.

\bibitem{gazette-times_police_nodate}
Lillian Schrock.
\newblock Police arrest person suspected of threatening a shooting at {OSU}.
\newblock {\em Corvallis Gazette Times}, February 2018.

\bibitem{shane_bergsma_broadly_2013}
{Shane Bergsma}, {Mark Dredze}, {Benjamin Van Durme}, {Theresa Wilson}, and
  {David Yarowsky}.
\newblock Broadly {Improving} {User} {Classification} via
  {Communication}-{Based} {Name} and {Location} {Clustering} on {Twitter}.
\newblock In {\em Proc. of the 2013 {North} {American} {Chapter} of the
  {Association} for {Computational} {Linguistics}: {Human} {Language}
  {Technologies} ({Hlt}-{NAACL})}, pages 1010--1019, Atlanta, GA, 2013. ACL.

\bibitem{sloan_who_2015}
Luke Sloan and Jeffrey Morgan.
\newblock Who {Tweets} with {Their} {Location}? {Understanding} the
  {Relationship} between {Demographic} {Characteristics} and the {Use} of
  {Geoservices} and {Geotagging} on {Twitter}.
\newblock {\em PLOS ONE}, 10(11):e0142209, November 2015.

\bibitem{sloan_knowing_2013}
Luke Sloan, Jeffrey Morgan, William Housley, Matthew Williams, Adam Edwards,
  Pete Burnap, and Omer Rana.
\newblock Knowing the {Tweeters}: {Deriving} {Sociologically} {Relevant}
  {Demographics} from {Twitter}.
\newblock {\em Sociological Research Online}, 18(3):1--11, August 2013.

\bibitem{sweeney_simple_2000}
Latanya Sweeney.
\newblock Simple {Demographics} {Often} {Identify} {People} {Uniquely}.
\newblock Working paper~3, Carnegie Mellon University, Laboratory for
  International Data Privacy, Pittsburgh, PA, 2000.

\bibitem{the_university_of_chicago_crime_lab_connect_2019}
{The University of Chicago Crime Lab}.
\newblock Connect \& {Redirect} to {Respect}: {Final} {Report}, January 2019.

\bibitem{tyler_mccormick_using_2015}
{Tyler McCormick}, {Hedwig Lee}, {Nina Cesare}, {Ali Shojaie}, and {Emma
  Spiro}.
\newblock Using {Twitter} for {Demographic} and {Social} {Science} {Research}:
  {Tools} for {Data} {Collection} and {Processing}.
\newblock {\em Sociological Methods and Research}, 46(3):390--421, 2015.

\bibitem{Western2006}
Bruce Western.
\newblock {\em Punishment and {{Inequality}} in {{America}}}.
\newblock {Russell Sage Foundation}, {New York}, 2006.

\end{thebibliography}

\newpage
\appendix
\section{Inter-rater reliability}
We established the reliability of the coding protocol using a sub-sample of 99 accounts from the randomly ordered, mixed data set. Three coders (the authors and an undergraduate research assistant) applied the protocol to this sub-sample.

We measure the reliability of our coding protocol by measuring the inter-rater reliability of our three coders using Krippendorff's alpha. Krippendorff's $\alpha = 1-D_o/D_e$ where $D_o$ is the observed disagreement between the coders and $D_e$ is the disagreement that would be expected by chance. We evaluated the resulting reliability measures using the benchmarking method proposed by Gwet\cite{kilem_gwet_handbook_2014} (as implemented in Stata by Klein\cite{daniel_klein_kappaetc:_2016}), indicating which of the Landis and Koch\cite{landis_measurement_1977} levels of agreement are met with probability at least 95\%.  As indicated in Table\ref{tab:rel}, our coders achieved at least a substantial level of agreement on the Landis and Koch scale for all our coding dimensions.

The levels of inter-rater reliability are reported in Table\ref{tab:rel}.  In coding Twitter accounts, the coders used a single label with three choices: \{Not accessible (protected, deleted, or suspended), Individual, Organization\}.  The first of these three choices can be done programmatically, but as we have observed, accounts become inaccessible or accessible over time as accounts are suspended and reinstated or made protected over time.  Since coding was not completed simultaneously, we opted to include ``Not accessible'' in the first feature, with ``Individual'' and ``Organization''.  However, the choices for this first feature impact our measure of inter-rater reliability for gender and race/ethnicity.  When measuring inter-rater reliability for gender and race/ethnicity, we included all subjects that had at least two coders select ``Individual'' for the first feature.  In order to understand the reliability of coding for race and ethnicity, in addition to considering the full set of 7 categories, we also considered a collapsed set of categories consisting of \{White, Not White, Don't Know\} where ``Not White'' consists of all remaining race and ethnicity categories.

\begin{table}[ht]\centering
  \begin{tabularx}{\columnwidth}{l *{4}{c}} \toprule
    & Indiv? & Gender & \multicolumn{2}{c}{Race \& Ethnicity} \\ \cmidrule(lr){2-2}\cmidrule(lr){3-3}\cmidrule(lr){4-5}
    \# categories & 3 & 4 & 7 & 3 \\
    \# subjects & 99 & 70 & 70 & 70 \\
    $\alpha$ & 0.940 & 0.884 & 0.703 & 0.785 \\ \midrule
    reliability & \multicolumn{2}{c}{almost perfect} & \multicolumn{2}{c}{substantial}\\\bottomrule
\end{tabularx}
  \caption{Inter-rater reliability}
  \label{tab:rel}
\end{table}
With the inter-rater reliability established, one of the three test coders (an undergraduate research assistant), coded the full, randomized, mixed data set. These codes serve as the basis for the analyses below.
\newpage
\section{Full Race and Ethnicity Demographics} \label{sec:demo-race-all}
\begin{table} [ht]\centering
  \begin{tabular}{lrrrr}\toprule
      & \makecell[c]{Corvallis} \\
      &\makecell[c]{Geotagging} & \multicolumn{3}{c}{DigitalStakeout} \\ \cmidrule(lr){3-5}
&	\makecell[c]{Tweeters}&	\makecell[c]{Narc.}&	\makecell[c]{Terr.}&	\makecell[c]{N+T}\\\midrule
$n$&	\makecell[c]{788}&	\makecell[c]{148}&	\makecell[c]{47}&	\makecell[c]{180}\\\midrule
White&	71.8\% &	78.4\% &	83.0\% &	78.9\%  \\
Black &6.5\% &	7.4\% &	4.3\% &	7.2\%  \\
Native American&	-&	-&	-&	-\\
Asian&2.8\% &	1.4\% &	- &	1.1\% 	\\
Native HI/Pac.&	0.4\% &	0.7\% &	- &	0.6\% \\
Middle Eastern	&4.6\% &	2.7\% &	- &	2.2\% \\
Other&	2.3\% &	2.0\% &	2.1\% &	2.2\% \\
    Hispanic&11.7\% &	7.4\% &	10.6\% &	7.8\% 	\\ \bottomrule
  \end{tabular}
  \caption{Coded Demographics: Narcotics \& Terror Report}
\end{table}

\begin{table} [ht]\centering
  \begin{tabular}{lr}\toprule
    $n$ & 242\\ \midrule
White&	84.3\% \\
Black &4.1\% \\
Native American&	-\\
Asian&2.1\% \\
Native HI/Pac.&	- \\
Middle Eastern	&1.2\% \\
Other&	2.5\% \\
    Hispanic&5.8\% \\\midrule
    Total	&100.00\%\\\bottomrule
  \end{tabular}
  \caption{Coded Demographics: LE Terms}
\end{table}

\newpage
\section{Demographics: Gender}\label{sec:demo-gender}

In Table\ref{tab:ds-cvi-twitter-gender}, we report on the gender ratios of Twitter users in our various data sets as described in the previous section.  Across the entire data set, CT and DS, only three users were coded ``Other''.  We removed these users from Table\ref{tab:ds-cvi-twitter-gender}.
We did pairwise comparisons between Narc and CVI and Terr and CVI, but not between LE and CVI as described in Section\ref{sec:demog}, with Chi-squared significance values in Table\ref{tab:ds-cvi-twitter-gender}.  As with race and ethnicity, we find that users in the DigitalStakeout data are representative of the Corvallis Geotagging Twitter users.

\begin{table}[ht] \centering
 \begin{tabular}{lrrrr}\toprule
      & \makecell[c]{Corvallis} \\
      &\makecell[c]{Geotagging} & \multicolumn{3}{c}{DigitalStakeout} \\ \cmidrule(lr){3-5}
&	\makecell[c]{Tweeters}&	\makecell[c]{Narc.}&	\makecell[c]{Terr.}&	\makecell[c]{LE}\\\midrule
   $n$&	\makecell[c]{788}&	\makecell[c]{148}&	\makecell[c]{47}&	\makecell[c]{241}\\\midrule
   M & 47.5\% &53.0\% &	47.8\% & 56.8\%\\
   F & 52.5\% & 47.0\% &	52.2\%& 43.2\%\\ \midrule
   $p$ & - & 0.25 & 1 & - \\\bottomrule
  \end{tabular}   \\ (LE collected in a different way from Corvallis Geotagging Tweeters, Narc., and Terr.)
  \caption{Coded Gender: Twitter Users}
  \label{tab:ds-cvi-twitter-gender}
\end{table}

In Table\ref{tab:census-gender} we give the gender distribution of the residents of Corvallis, OR and of Oregon State University students.  Chi-squared tests indicate that DigitalStakeout data may be representative of Corvallis residents ($p = 0.11$), but are not
representative of Oregon State University students ($p = 0.0022$).  The latter comparison may be more valid given that Twitter use is more prevalent among the young (student-aged), but OSU's demographics also include those of online students who may not Tweet from Corvallis.

\begin{table} [ht]\centering
  \begin{tabularx}{0.6\columnwidth}{lcc}\toprule
& \makecell[c]{OSU}&\makecell[c]{Corvallis}\\\midrule
    M   & 52.9\%& 50.3\%
\\
    F & 47.1\%  & 49.7\%
\\\bottomrule

\end{tabularx}
  \caption{Gender distribution: Oregon State University and Corvallis (Census)}
  \label{tab:census-gender}
\end{table}
\newpage
\section{Keyword roots and their variants}
\begin{table}[ht]
  \centering
  \caption{Necessary keyword roots and the variants that appear in DigitalStakeout Tweets.}
  \begin{tabular}[t]{ll}
    Root & Variants \\ \toprule
    snow	&	\#snow,  snow,  \#crouchingtigerhiddensnowman,  \\ & \#snowboard,  \#snowday,  \#snowinmarch	\\
hop	&	hop,  hopping,  hops,  hopped,  \#hops	\\
high	&	high,  highland,  \#highcbd,  skyhigh	\\
line	&	line,  lining	\\
party	&	party,  \#monkeysparty,  \#kidsparty,  \#party,   \\ & \#pizzaparty,  \#birthdayparty,  \#partyfoul	\\
smoke	&	smoke,  smoking,  smoked	\\
bowl	&	bowl,  bowling,  bowls	\\
rock	&	rock,  rocked,  rocking,  rocks	\\
face	&	face, facing, faces, faced	\\
cheese	&	cheese, cheesy	\\
bag	&	bag, bags	\\
jack	&	jack, jacked	\\
treat	&	treat, treats	\\
blast	&	blast, blasted	\\
fried	&	fried, fries	\\
crystal	&	crystal, crystals	\\
trip	&	tripped, tripping, trips	\\
burger	&	burger, burgers	\\
cook	&	cooked, cooking	\\
dope	&	dope, \#dopeman	\\
blow	&	blowing, blow, blower	\\
wreck	&	wrecking, wreck	\\
bake	&	baking, bake	\\
waste	&	wasted, wasting	\\
gang	&	gang, \#canongang	\\
hustle	&	hustled, hustle	\\
rip	&	rip, \#ripmicrophone	\\
  \end{tabular}
\end{table}

\begin{table}[ht]
  \centering
  \caption{Likely keyword roots and the variants that appear in DigitalStakeout Tweets.}
  \begin{tabular}[t]{ll}
    Root & Variants \\ \toprule
    pie	&	pie, \#pieeatingcontest	\\
pot	&	pot, \#pot, \#pothead, \#potfarm	\\
zone	&	zone, calzone	\\
bud	&	bud, budtender, buds, \#budtenders	\\
fade	&	fades, faded, \#functionfades, \#faded	\\
dabpro	&	\#thedablab, dabpro	\\
bang	&	bang, banged	\\
deal	&	deal, dealers, deals, dealt	\\
indica	&	\#indica, indica	\\
mash	&	mash, mashed, mashing	\\
dank	&	dank, \#danksgiving	\\
hip	&	hip, hippie, hipster	\\
jam	&	jam, \#ujam, jamming	\\
angel	&	angel, angeles	\\
addict	&	addicts, addict, \#slapaddictz	\\
roller	&	roller, \#rollpipps, rolled	\\
\#growweed	&	\#growweed, \#weedagram	\\
burn	&	burn, burning	\\
keg	&	kegs, keg	\\
malt	&	malt, malts	\\
melt	&	melt, melting	\\
  \end{tabular}
\end{table}
\begin{table}[ht]
  \centering
  \caption{Known keyword roots and the variants that appear in DigitalStakeout metadata.}
  \begin{tabular}[ht]{ll}
    Root & Variants  \\ \toprule
    angel	&	angel, 	angeles	\\
blast	&	blast	, 	blasted	\\
blow	&	blow, 	blowing, blower	\\
bowl	&	bowl, 	bowling,  bowls	\\
broken	&	broken, 		\\
burn	&	burned, 	burn,  burning	\\
cheese	&	cheese, 	cheesy	\\
cook	&	cooked, 	cooking	\\
crystal	&	crystal, 	crystals	\\
dope	&	dope, 	\#dopeman	\\
face	&	faced, 	face, facing, faces	\\
fade	&	faded, 	fades,  \#functionfades, \#faded	\\
fried	&	fried, 	fries	\\
high	&	high, 	highland, \#highcbd, skyhigh, highness	\\
hookup	&	hookup, 		\\
hop	&	hopped, 	hop,  hops,  hopping,  \#hops	\\
indica	&	indica, 	\#indica	\\
line	&	line, 	lining	\\
load	&	loaded, 	loading	\\
mash	&	mashed, 	mash, mashing	\\
munchies	&	munchies, 	munchys	\\
party	&	party, 	\#monkeysparty, \#kidsparty, \#party, \\& \#pizzaparty, \#birthdayparty, \#partyfoul	\\
pied	&	pied, 	pie, \#pieeatingcontest	\\
pot	&	pot, 	\#pot, \#pothead, \#potfarm	\\
rock	&	rock, 	rocked, rocking, rocks	\\
smoke	&	smoke, 	smoking, smoked	\\
snow	&	snow, 	\#snow, \#crouchingtigerhiddensnowman, \\& \#snowboard, \#snowday, \#snowinmarch	\\
stash	&	stash, 		\\
stuck	&	stuck, 		\\
track	&	tracks, 	track, tracked	\\
trip	&	trip, 	tripped, tripping, trips	\\
tweed	&	tweed, 		\\
waste	&	wasted, 	wasting	\\
wreck	&	wreck, 	wrecking,  wrecked	\\
yay	&	yay, 		\\
zone	&	zoned, 	zone, calzone
  \end{tabular}
\end{table}
\end{document}